\documentclass[aps,pra,amsmath,amssymb,reprint,superscriptaddress,floatfix]{revtex4-2}

\usepackage{graphicx}% Include figure files
\usepackage{amsmath}
\usepackage{amssymb}
\usepackage{float}
\usepackage{bm}% bold math
\usepackage[T1]{fontenc}
\usepackage[latin9]{inputenc}
\usepackage{nicefrac, xfrac}
\usepackage[unicode=true,pdfusetitle,
 bookmarks=true,bookmarksnumbered=false,bookmarksopen=false,
 breaklinks=false,pdfborder={0 0 1},backref=false,colorlinks=false]
 {hyperref}
\hypersetup{
 colorlinks,linkcolor=red,anchorcolor=black,citecolor=blue,breaklinks=true}
\makeatletter
\@ifundefined{textcolor}{}
{%
 \definecolor{BLACK}{gray}{0}
 \definecolor{WHITE}{gray}{1}
 \definecolor{RED}{rgb}{1,0,0}
 \definecolor{GREEN}{rgb}{0,1,0}
 \definecolor{BLUE}{rgb}{0,0,1}
 \definecolor{CYAN}{cmyk}{1,0,0,0}
 \definecolor{MAGENTA}{cmyk}{0,1,0,0}
 \definecolor{YELLOW}{cmyk}{0,0,1,0}
}

\begin{document}

\title{Out-of-Time-Order-Correlator for van der Waals potential}% Force line breaks with \\

\author{Hui Li}
\author{Eli Halperin}
\author{Reuben R.W. Wang}
\author{John L. Bohn}
\email{hui.li@jila.colorado.edu}
\affiliation{Joint Institute for Laboratory Astrophysics, University of Colorado, Boulder, Colorado 80309, USA}


\date{\today}% It is always \today, today,
\begin{abstract}
    The quantum-to-classical correspondence is often quantified in dynamics by a quantity referred to as the out-of-time-order correlator (OTOC). In chaotic systems, the OTOC is expected to grow exponentially at early time, characteristic of  a Lyapunov exponent, however, exponential growth can also occur for integrable systems.  Here we investigate the OTOC for realistic diatomic molecular potentials in one degree of freedom, finding that the OTOC can grow exponentially  near the dissociation energy of the moelcule. Further, this dynamics is tied to the classical dynamics of the atoms at the outer classical turning point of the potential.  These results should serve to guide and interpret dynamical chaos in more complex molecules. 
\end{abstract}

\maketitle

%\tableofcontents
\section{Introduction}

The link between quantum mechanical and classical physics remains an elusive one.  Currently, one of the frontline efforts in forging this link seeks quantum mechanical quantities whose time evolution depends sensitively on an initial condition, reminiscent of the Lyapounov exponent in classical chaos \cite{rozenbaum2017, garcia2018, gharibyan2019, chavez2019, lewis2019, fortes2019,  ali2020, rautenberg2020, bhagat2020, xu2020, pilatowsky2020, wang2021, yin2021, bhattacharyya2022, zhang2022}.  One such quantity, and the one we will consider here, is the  out-of-time-order correlator, which we will, following a merciful convention, refer to simply by its acronym, OTOC \cite{larkin1969, hashimoto2017, chavez2019}.  We will motivate and define this quantity in our context below, but here we note that it is at least a partial success: for example, in a model problem such as a pair of nonlinearly coupled harmonic oscillators, the quantum mechanical OTOC grows exponentially in time under the same circumstances in which the classical system exhibits positive Lyapunov exponents \cite{akutagawa2020}.

Not all mechanical systems are chaotic, however.  In particular, non-dissipative motion in a single degree of freedom, with a time-independent Hamiltonian, is always integrable (having at least total energy as a constant of the motion). The OTOC has been studied in various such instances, finding that it does indeed show exponential growth in certain situations, such as an inverted harmonic oscillator potential, where classical trajectories also depend exponentially sensitively on initial conditions \cite{hashimoto2020, ali2020, qu2022, morita2022}.  If not ``chaotic'', let us at least refer to this kind of behavior as ``sensitive''. Vice versa, in restoring potentials such as a regular harmonic oscillator \cite{hashimoto2017}, or a power law potential $\propto x^{2N}$ \cite{romatschke2021}, exponential growth is seen in neither the classical dynamics nor in the OTOC, a situation that might be termed ``regular''.  Thus it appears that the OTOC may serve as a quantum-to-classical link in dynamics, irrespective of the chaotic nature of that dynamics.

In this paper we extend the work on OTOCs in one-dimensional systems, to potentials that exist between atoms in a molecule, for example, a schematic Lennard-Jones potential, as well as the realistic singlet potential between ground state rubidium atoms. Such potentials present a unique opportunity to explore the behavior of the OTOC, since:
\begin{enumerate}
\item The low-lying energy eigenstates are well-represented by harmonic oscillators, hence show no exponential sensitivity to either classical initial conditions, or to the OTOC.
\item By contrast, higher-lying, very anharmonic states can show exponential dependence in both these quantities. Thus a transition from regular to sensitive behavior can be described.
\item Wave functions of the high-lying states are strongly concentrated at the outer classical turning point $r_c$ of the potential.  As we will see, this emphasizes the essentially local character of the quantum-to-classical link, at least at short times, and its ultimate origin in the sign of the curvature of the potential near $r_c$.
\end{enumerate}

Ruminations of this sort are of interest beyond a single degree of freedom. For example, it has been established experimentally that weakly-bound dimers of lanthanide atoms, explored in ultracold gases, show hints of quantum chaos via a standard signature, namely,  their nearest-neighbor level spacings exhibit  statistics close to that implied by the Gaussian Orthogonal Ensemble in random matrix theory  \cite{frisch2014, maier2015}. This kind of chaos is expected to be driven by strong nonlinear couplings between the ro-vibrational and spin degrees of freedom of these highly multichannel molecules \cite{makrides2018, mccann2021}.  Demonstrable chaos in the molecules' dynamics has, however, yet to be addressed, either theoretically or experimentally.  It is conceivable that the OTOC would be the means to do so, but for this approach to be viable, we must first understand how the OTOC works in a single molecular channel, so that mere sensitivity will not be mistaken for truly chaotic behavior. It is also worth  noting that exponentially growing OTOCs also correlate to classical Lyapunov exponents in the spectra of polyatomic molecules \cite{zhang2022}.

The remainder of this work is structured as follows: After a brief motivation in Sec.~\ref{sec:md}, we describe the model systems used and the numerical methods to calculate the OTOC in Sec.~\ref{sec:mcm}. Numerical results are presented in Sec.~\ref{sec:rf}, and in Sec.~\ref{sc:int} we detail an approximation to deduce the dependence of the exponential growth behavior. Finally, we draw our conclusions with a discussion on the further possible research directions in Sec.~\ref{sc:co}.

\section{Motivation and Definition}
\label{sec:md}

The quantum mechanical OTOC is defined by a close analogy with classical dynamics. In one degree of freedom this is formulated as follows. For a generic classical Hamiltonian
\begin{align}
H = \frac{ p^2 }{ 2m }  + V(x),
\end{align}
Hamilton-Jacobi theory posits the ability to transform between alternative representations of coordinate and momentum \cite{bohn2018}. In particular, a set of phase space coordinates may be described by the initial position and momentum, $(x_0,p_0)$, and related to the time-evolving coordinates, $(x(t),p(t))$, by an appropriate time-dependent canonical transformation \cite{calkin1996}.  The time-evolving coordinates are then explicit functions of the initial conditions, which in our immediate case can be written
\begin{align}
x(t) = x_0 f(t) + \frac{ p_0 }{ m \chi } g(t).
\end{align}
Here, $f(t)$ and $g(t)$ are solutions to the equations of motion, with initial conditions $f(0)=1$, ${\dot f}(0) = 0$, $g(0)=0$, ${\dot g}(0)=\chi$, and $\chi$ is a constant carrying units of frequency.  

Written this way, the relation between $x$ at some time $t$, and its initial value $x_0$, is fairly explicit. This relation is brought out by computing the Poisson bracket,
\begin{align}
\{ x(t), p_0 \}_{x_0,p_0} &= \frac{ \partial x(t) }{ \partial x_0 }\frac{ \partial p_0 }{ \partial p_0 }
- \frac{ \partial x(t) }{ \partial p_0 } \frac{ \partial p_0 }{ \partial x_0 } \nonumber \\
&=  \frac{ \partial x(t) }{ \partial x_0 }.
\label{eq:Poisson}
\end{align}
which therefore denotes the dependence of $x(t)$ on the initial condition $x_0$. Thus for a harmonic oscillator with $V(x) = m \omega^2 x^2/2$,  $\partial x(t) / \partial x_0 = \cos \omega t$, the dependence is regular rather than sensitive; whereas for an anti-oscillator, $V(x) = - m \lambda^2 x^2/2$, $\partial x(t) / \partial x_0 = \cosh \lambda t$ soon evolving to $ \sim e^{\lambda t}$, showing that the dependence is exponentially sensitive, where $\lambda$ is analogous to the frequency scale $\omega$ in the harmonic oscillator. Note that classical sensitivity in these two cases is tied to the sign of the potential' s curvature.

Thus motivated, the quantity of interest is transported to the quantum realm in the conventional way, by replacing the Poisson bracket by a commutator, $i.e.$, quantization, and by presenting operators in the Heisenberg representation. That is to say: given the quantum mechanical Hamiltonian
\begin{align}
{\hat H} = \frac{ {\hat p}^2 }{ 2m } + {\hat V}({\hat x}),
\end{align}
and the Heisenberg-representation coordinate operator
\begin{align}
{\hat x}(t) = e^{i {\hat H} t / \hbar} {\hat x}(0) e^{- i {\hat H} t / \hbar},
\end{align}
sensitivity to initial condition would be expressed by the quantity
\begin{align}
\left[ {\hat x}(t), \hat{p}(0) \right]/i\hbar.
\end{align}
which is the quantum analog of Eq.~(\ref{eq:Poisson}). Two addenda are made to this procedure.  First, one squares the commutator, lest its interesting parts average out; and second, one takes its average over a quantum state of interest; for us, this will be an energy eigenstate $|n \rangle$. 

Thus, the OTOC for state $n$ is defined as
\begin{align}
C_n(t) = -\langle n | \left[ {\hat x}(t), \hat{p}(0) \right]^2 |n \rangle/\hbar^2.
\label{eq:cnt1}
\end{align}
This is a four-point correlation function evaluated between operators with the times appearing out of order, hence the name ``out-of-time-order correlator''. The quantity $C_n(t)$ therefore defines a time-dependent number, allowing us to ascertain, for each $n$, whether $C_n(t)$ grows exponentially in time.  More generally, one sometimes computes a thermal average of the $C_n (t)$'s, but we do not do so here. Using the classical-quantum correspondence $[,]/i\hbar \rightarrow \left \{,\right\}$, then
\begin{align}
    \frac{-\langle n | \left[ {\hat x}(t), \hat{p}(0) \right]^2 |n \rangle}{\hbar^2} \rightarrow \{ x(t), p_0 \}^2 = \left ( \frac{ \partial x(t) }{ \partial x_0 } \right)^2.
\end{align}
Thus, $C_n(t) = \cos^2(\omega t)$ for a harmonic oscillator \cite{hashimoto2017}, whereas for an anti-oscillator $C_n(t) = \cosh^2(\lambda t) \sim e^{2\lambda t}$ \cite{morita2022}, indicating that the quantum sensitivity to initial condition is twice that in the classical case.

Throughout the following sections, atomic units are used if not specified.

\begin{figure*}[ht]
       \centering
       \begin{minipage}{1\textwidth}
         \bf{Lennard-Jones}
       \end{minipage}
    \includegraphics[scale=0.2,trim=15 10 15 0,clip]{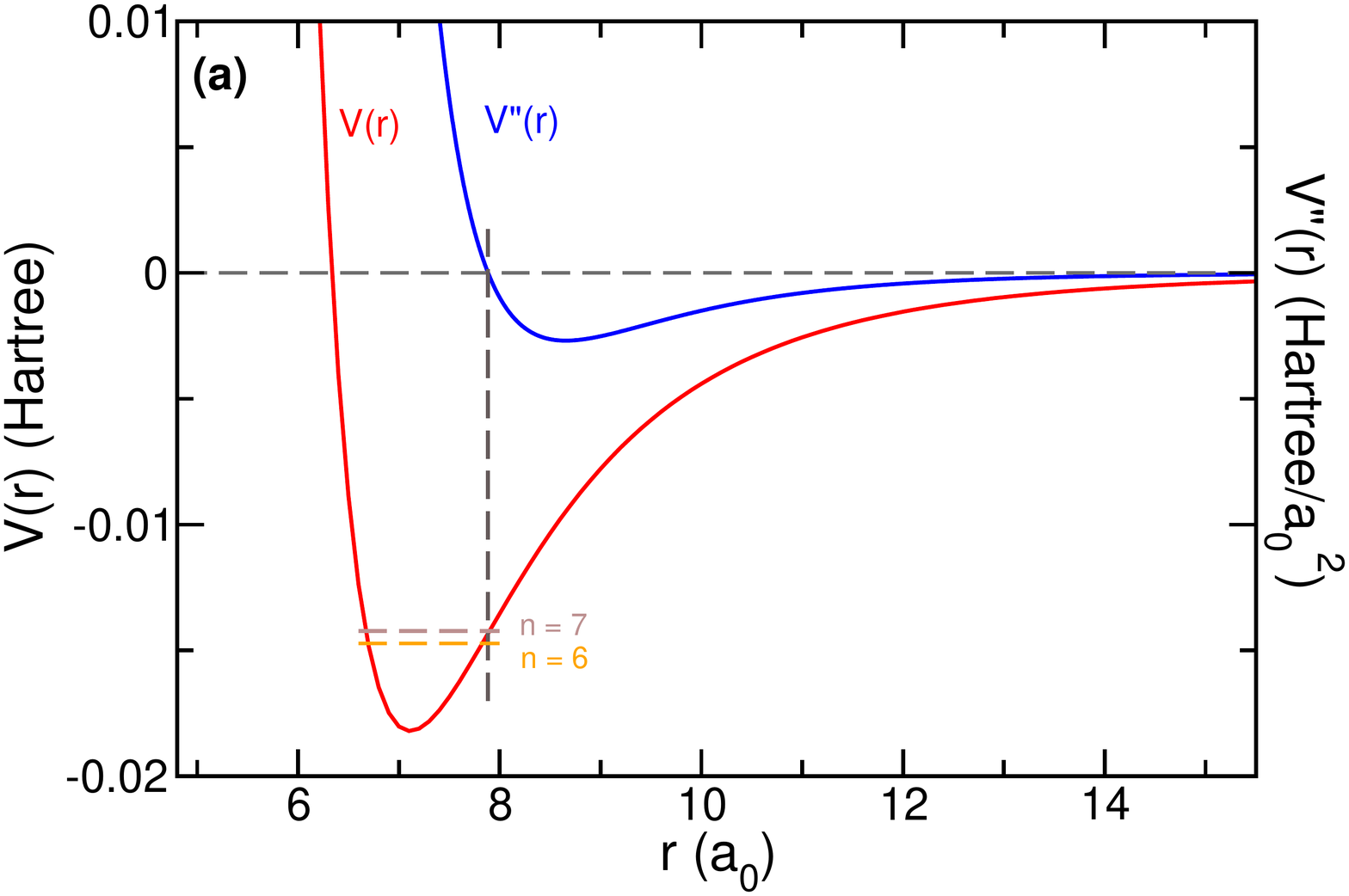}
    \includegraphics[scale=0.2,trim=15 10 15 0,clip]{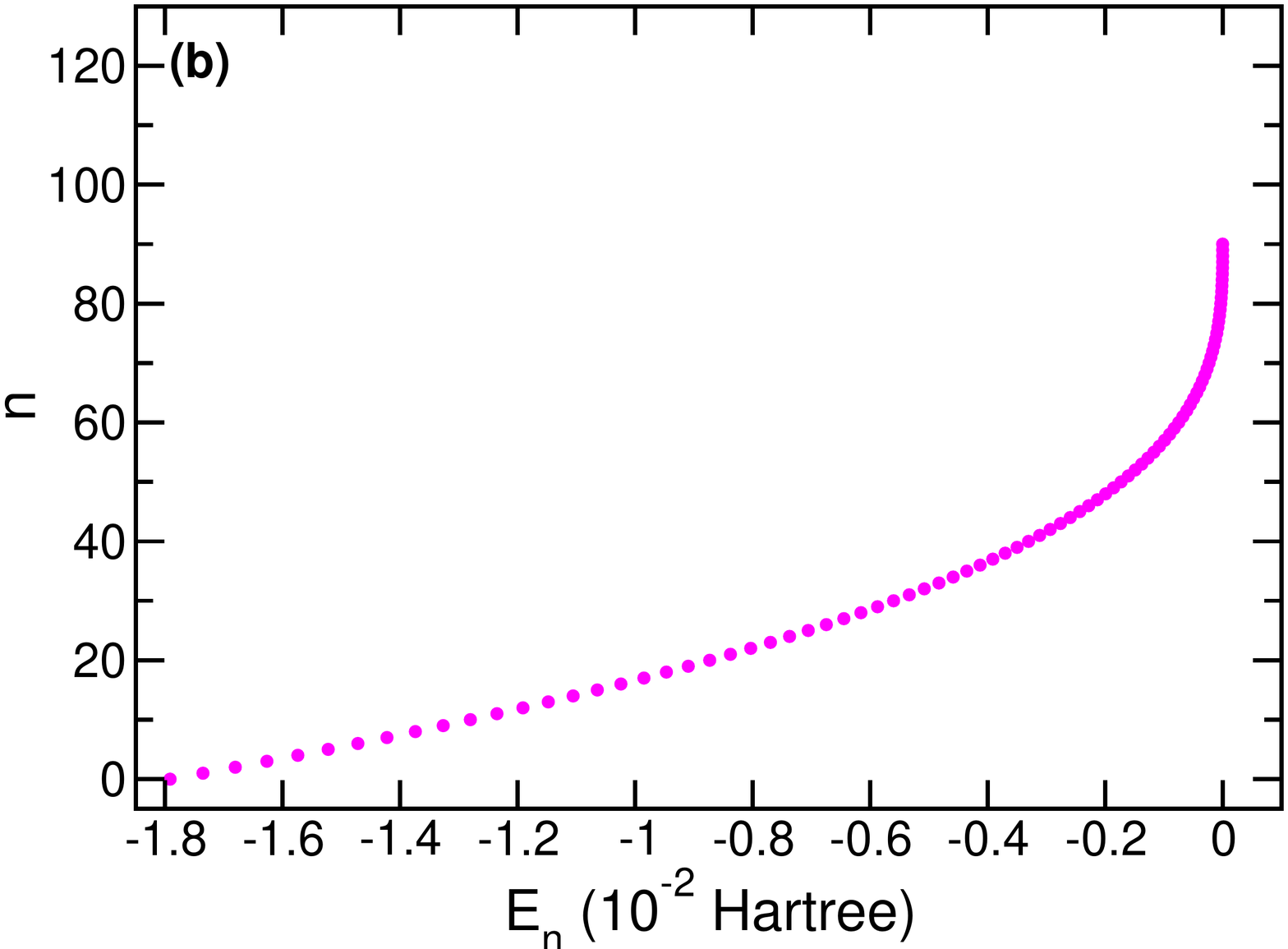}
    \includegraphics[scale=0.2,trim=15 10 15 0,clip]{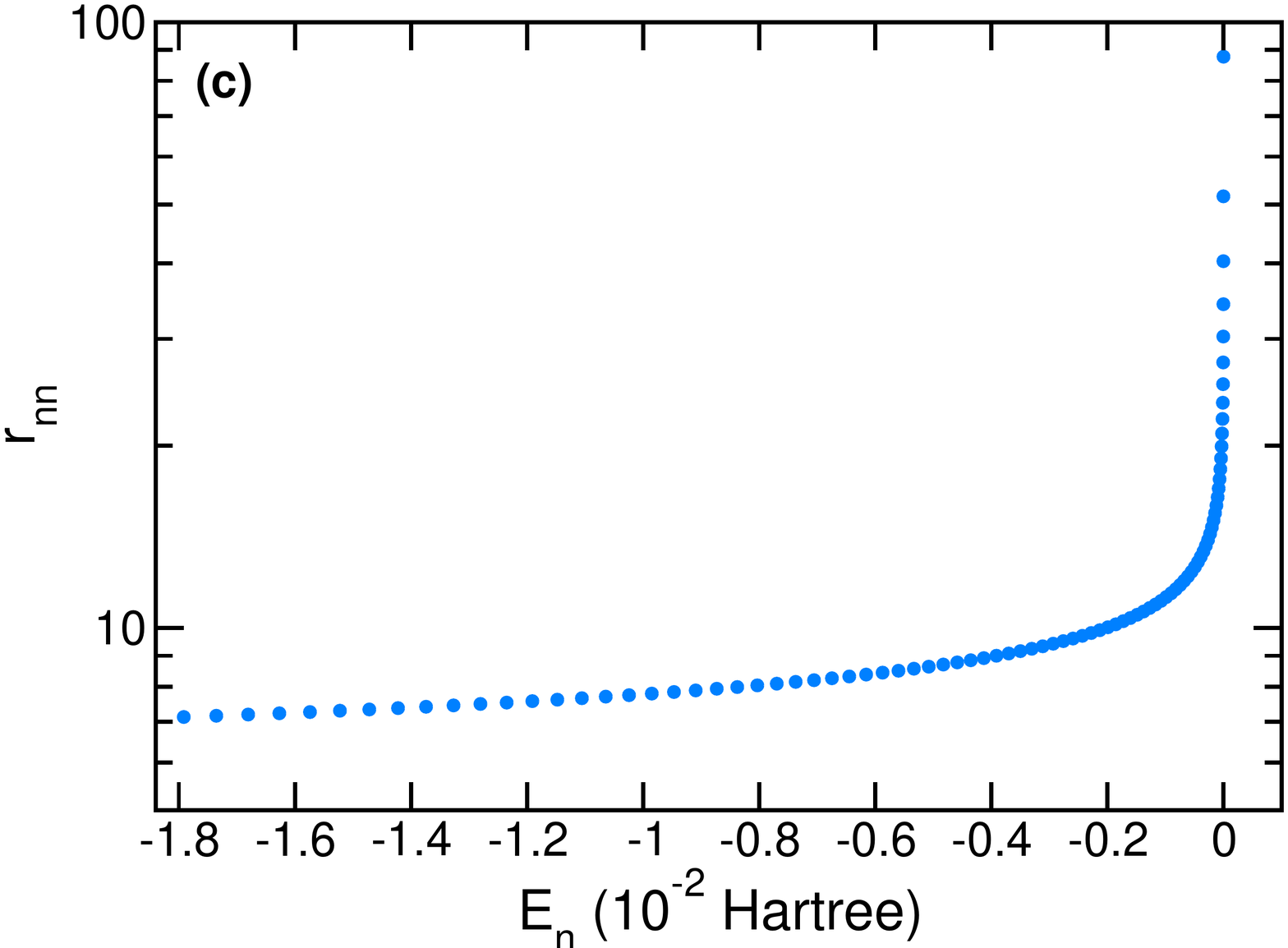}
     \begin{minipage}{1\textwidth}
         \vspace{10 pt}
         \bf{Rb$_2$ $X^{1}\Sigma^+_g$}
    \end{minipage}
    \includegraphics[scale=0.2,trim=15 10 15 0,clip]{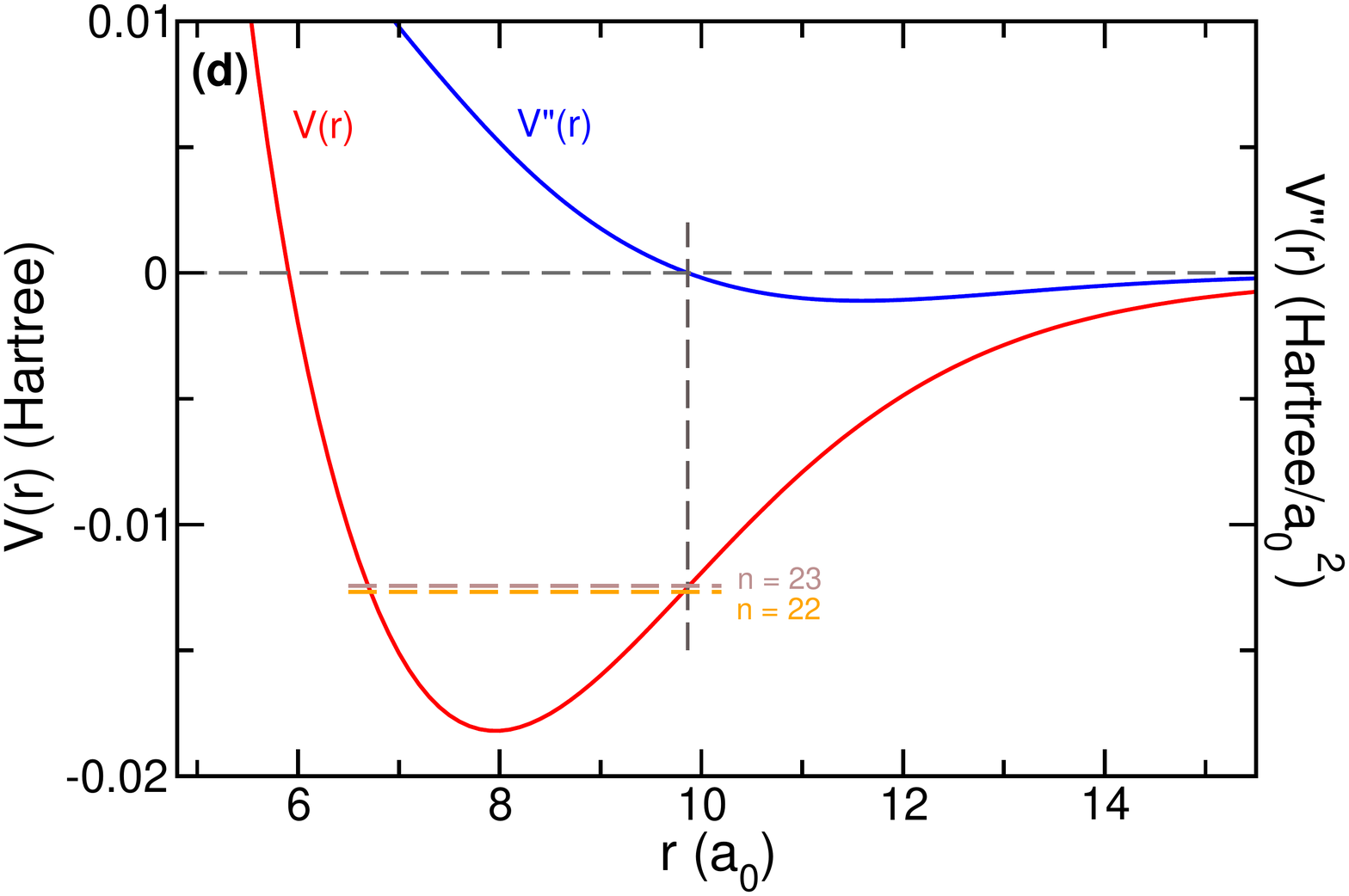}
    \includegraphics[scale=0.2,trim=15 10 15 0,clip]{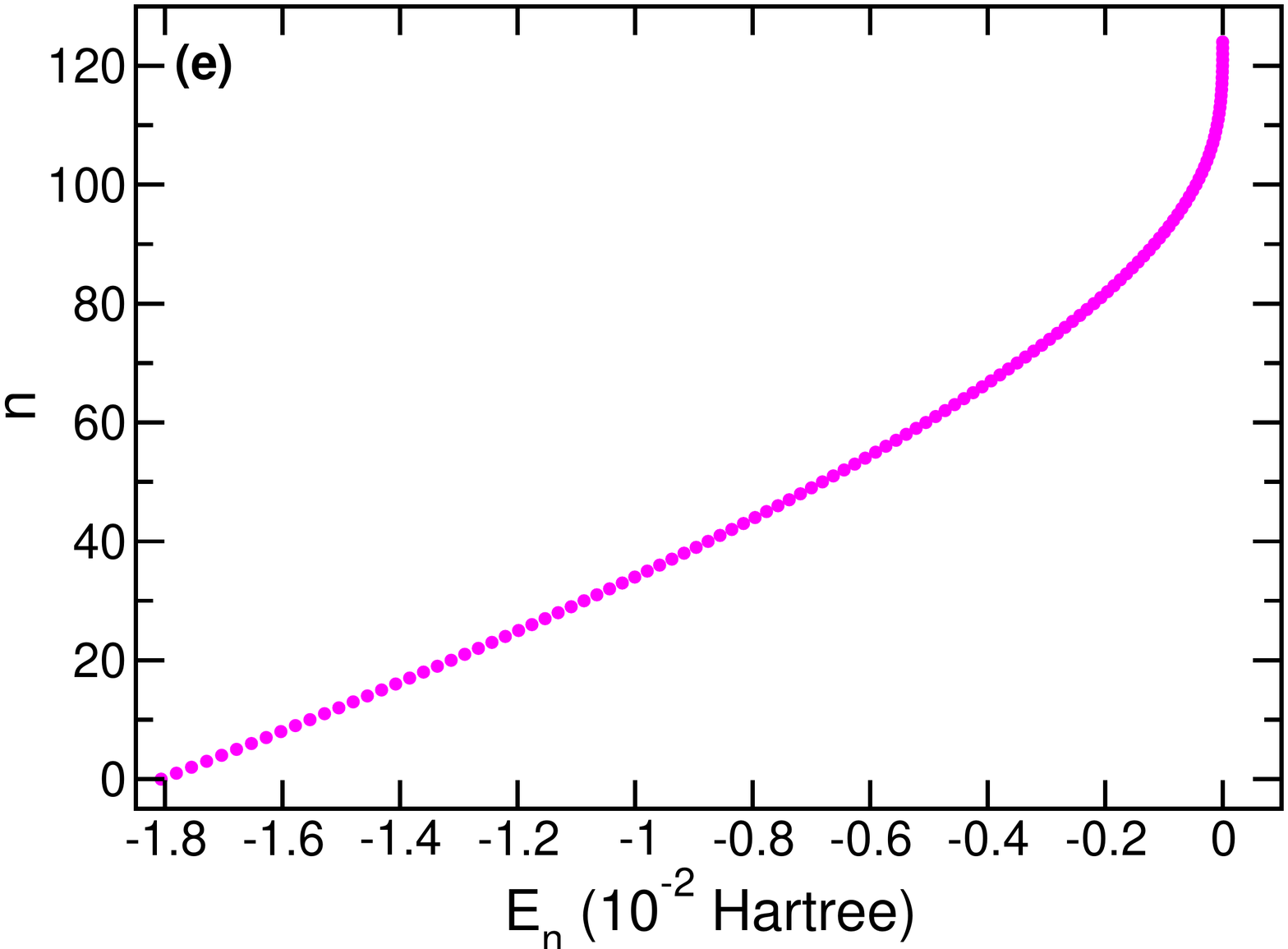}
    \includegraphics[scale=0.2,trim=15 10 15 0,clip]{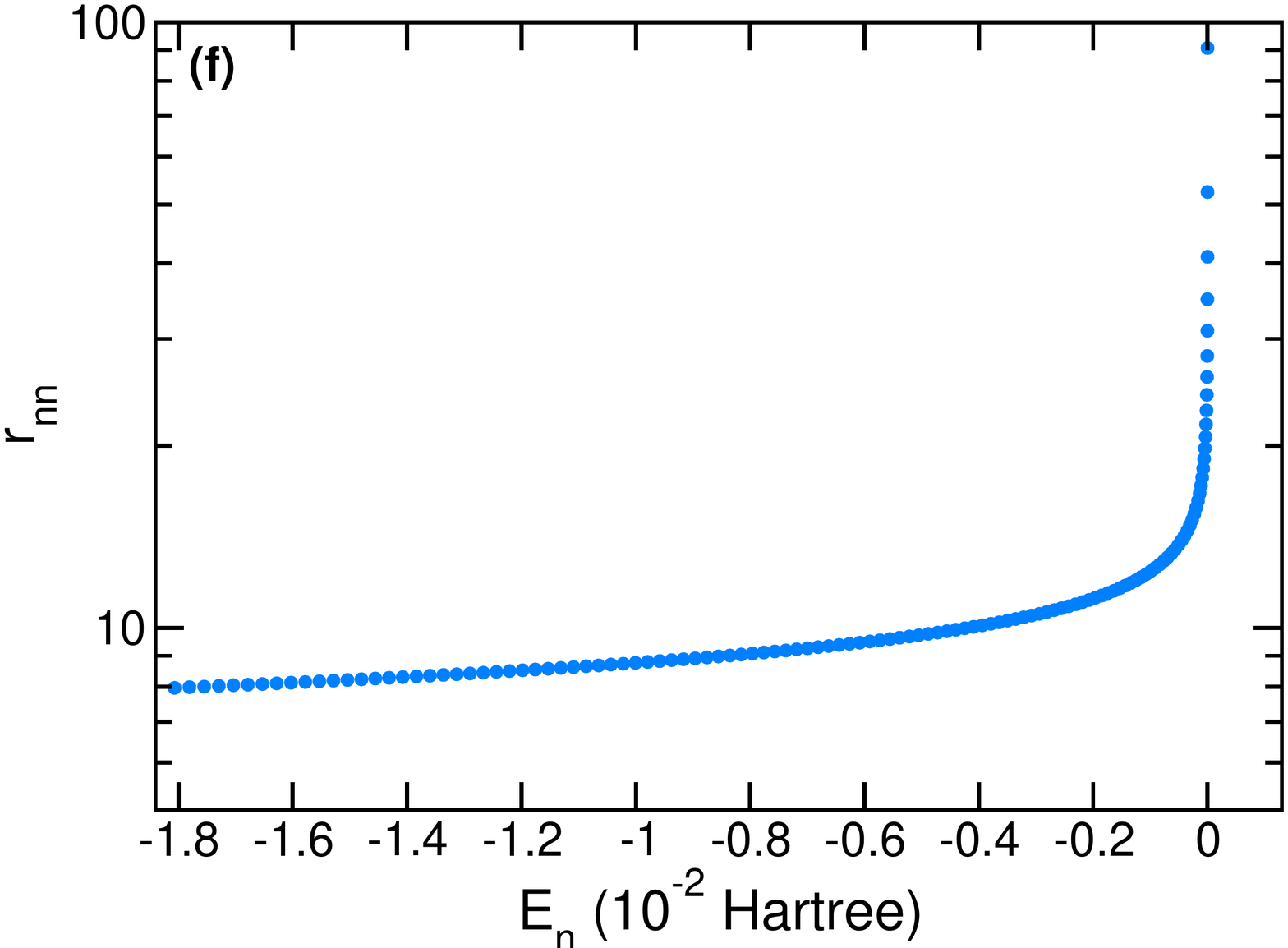}
    \caption{Upper panels (a), (b) and (c) show the potential (also the second order derivatives) as a function of interatomic separation, eigenstates spetra, and the matrix elements of operator $\hat{r}$ sorted by eigenvalues for the Lennard-Jones potential. Lower panels (d)-(f) present  $X^{1}\Sigma^+_g$ potential of two interacting rubidium atoms \cite{strauss2010}.}
    \label{fig:pot} 
\end{figure*}

\section{Model and Computing Methods}
\label{sec:mcm}

We consider two similar situations. The first consists of identical structureless atoms, interacting via the Lennard-Jones potential
\begin{align}
    V(r) = \frac{C_{12}}{r^{12}} - \frac{C_6}{r^6},
\end{align}
where $r$ is the interatomic separation. The second situation consists of two ground-state rubidium atoms, interacting via their $X^{1}\Sigma^+_g$ ground state potential \cite{strauss2010}. We keep the two situations similar, if not identical, by setting $C_6$ in the Lennard-Jones potential equal to the $C_6$ coefficient for rubidium; and by choosing $C_{12}$ so that the two potentials have the same depth. In both cases, we use the reduced mass $\mu$ of two rubidium atoms. To constrain the motion in single degree of freedom, we discard the molecular rotation in our model, $i.e.$, the atoms interact in a $s$-wave channel.

Denoting the basis of bound energy eigenstates of either potential by $|n\rangle$, the OTOC defined in Eq.~(\ref{eq:cnt1}) can be rewritten as,
\begin{align}
\label{eq:cnt2}
    C_n(t) = \sum_l b_{nl}(t)b^{*}_{nl}(t),
\end{align}
where
\begin{align}
    b_{nl}(t) &= -i\langle n|[ \hat{r}(t), \hat{p}(0) ]| l\rangle  \nonumber \\
             &= -i\sum_k \left ( e^{i E_{nk} t} r_{nk} p_{kl} - e^{i E_{kl} t} p_{nk} r_{kl} \right ),
             \label{eq:bnm}
\end{align}
here we denote $E_{nl} \equiv E_n - E_l$, $r_{nl} \equiv \langle n| \hat{r}(0) | l \rangle$ and $p_{nl} \equiv \langle n| \hat{p}(0) | l \rangle$. Since $[\hat{H}, \hat{r}(0)] = -i\hat{p}(0)/\mu$, then the matrix elements relation between the two operators can be derived, $p_{nl} = i\mu E_{nl}r_{nl}$. Substituting this relation into Eq.~(\ref{eq:bnm}), we have
\begin{align}
\label{eq:bnm2}
   b_{nl}(t) = \mu \sum_k r_{nk}r_{kl} \left ( E_{kl}e^{i E_{nk} t} - E_{nk}e^{i E_{kl} t} \right ).
\end{align}
This method provides with a concise framework for calculating the OTOC \cite{hashimoto2017}.

The eigenstate energy levels and wave functions are numerically solved by diagonalizing the exact Hamiltonian of the nuclear motion with a nonequidistant grid in the discrete variable representation (DVR) method \cite{tiesinga1998}. Note that computing the OTOC for eigenstate $|n\rangle$ requires a sum over eigenstates $|l\rangle$, $|k\rangle$ distinct from $|n\rangle$. For $|n\rangle$ too high in energy, this sum may require states above the dissociation threshold, which are not computed by the DVR method.  For this reason, in what follows we restrict $n$ to somewhat below the last bound state of the potentials, where we can assure convergence of $C_n (t)$ to approximately 1\%.

\section{Results}
\label{sec:rf}

\begin{figure*}
       \centering
       \textbf{Lennard-Jones \qquad\qquad\qquad\qquad\qquad\qquad\qquad\qquad} \textbf{Rb$_2$ $X^{1}\Sigma^+_g$}\par\medskip
    \includegraphics[scale=0.3,trim=15 10 15 0,clip]{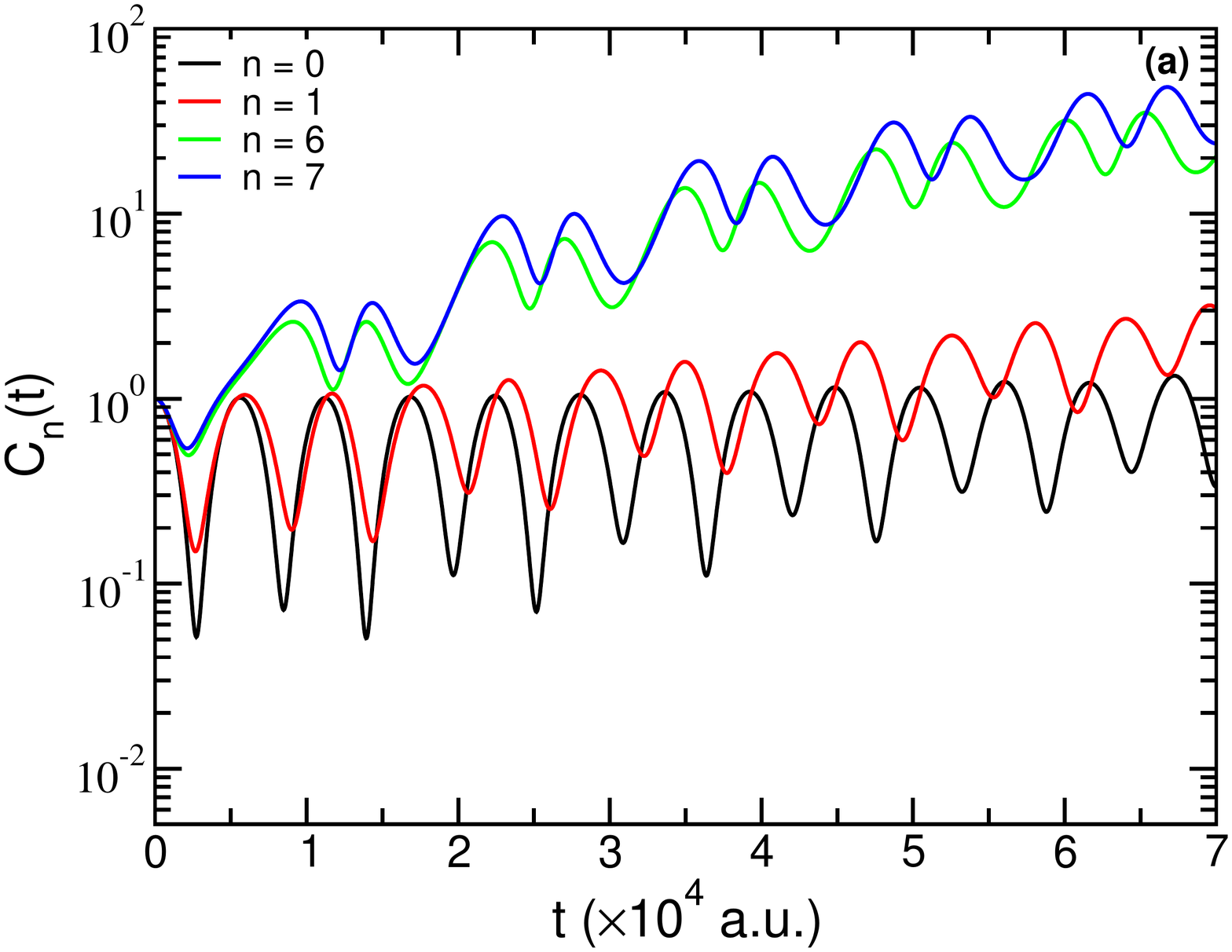}
    \includegraphics[scale=0.3,trim=15 10 15 0,clip]{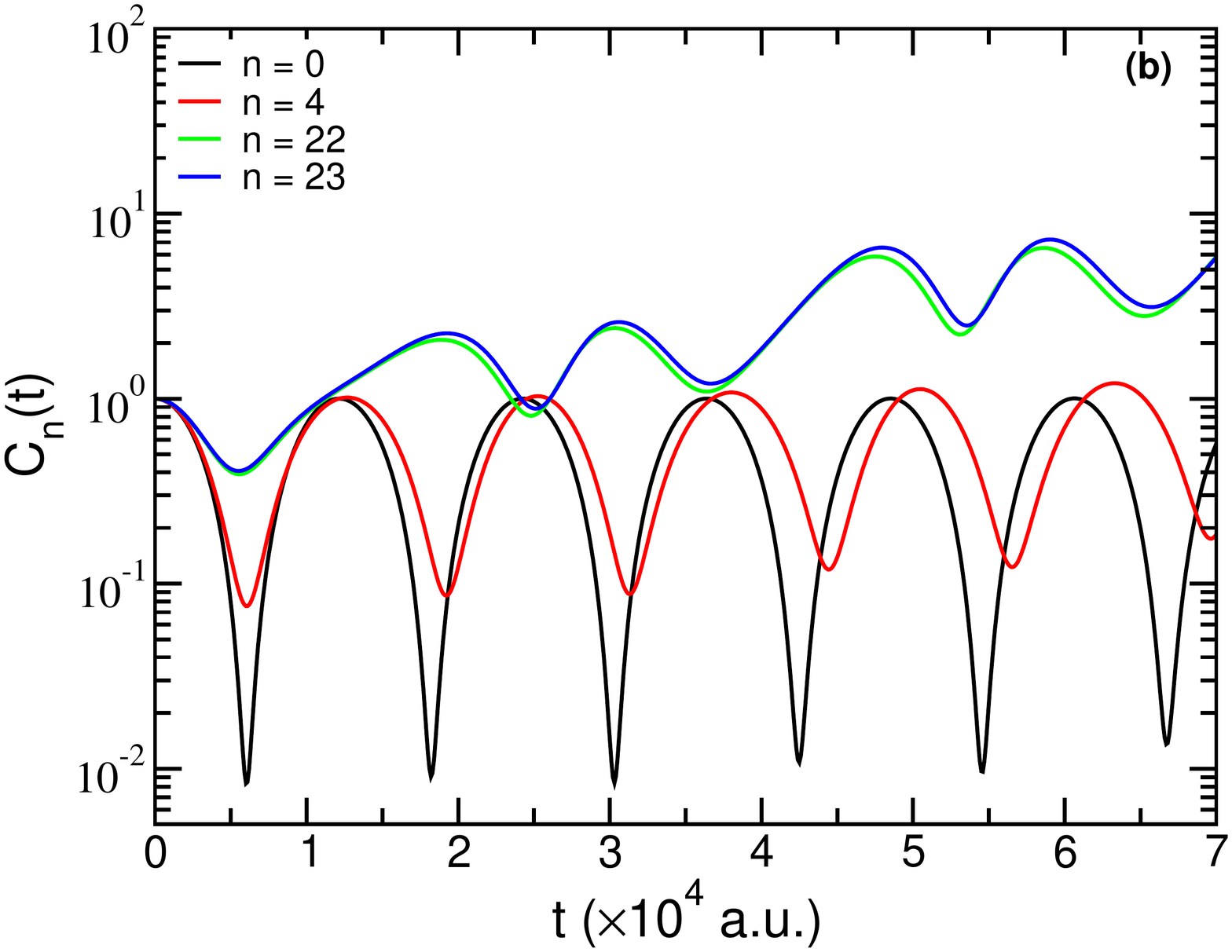}
    \caption{The OTOCs of lower energy modes for Lennard-Jones potential (a ), and Rb$_2$ $X^{1}\Sigma^+_g$ potential (b), respectively.}
    \label{fig:m1} 
\end{figure*}

\subsection{Potential and Eigenvalue spectrum}

The two potentials are shown in Figs.~\ref{fig:pot}(a) and (d) (red lines). The other panels depict the energy spectrum (b, e) and the mean interatomic separation $r_{nn}$ (c, f) of these potentials. We can see that even with the same well depth, the Lennard-Jones potential supports fewer bound-states than the realistic $X^{1}\Sigma^+_g$ potential due to its narrower potential shape.

A key component in describing the OTOC in a one-dimensional potential, as described above, is the curvature of the potential: regions of coordinate where this curvature is negative are expected to correlate roughly to exponential OTOC growth. For this reason, the second derivative of the potentials is also shown (blue) in Figures \ref{fig:pot}(a) and \ref{fig:pot}(d). Also noted is the range of $n$ where the potential at the classical outer turning point nearly coincides with the transition from positive to negative curvature. As shown, this point is between $n=6$ and $7$ for the Lennard-Jones potential, and between $n=22$ and $23$ for the realistic rubidium dimer potential. This is therefore the scale of $n$ on which the transition to exponentially growing OTOC may be expected to set in.

\subsection{OTOC Calculation}

Beginning with the low-energy part of the spectrum, Fig.~\ref{fig:m1}(a) shows the time evolution of the OTOC for selected energy eigenstates $|n\rangle$ of the Lennard-Jones potential. For the vibrational ground $n=0$ and first-excited ($n=1$) states, the OTOC merely oscillates in time, consistent with the bottom of this potential being approximately harmonic. The period of oscillation is about 5520 a.u., consistent with the equivalent harmonic oscillator. Slightly higher in the spectrum, for $n=6,7$, the OTOC experiences a brief burst of exponential-like growth up to $\approx 10^{-4}$ a.u., before again varying quasi-sinusoidally. Recall that this is the approximate value of $n$ at which exponential growth may be expected to start, but it has not yet taken over decisively.

Similarly, for the realistic rubidium $X^1\Sigma^+_g$ potential as shown in Fig.~\ref{fig:m1}(b), the OTOC for the low-lying states $n=0,4$ vary sinusoidally. Higher in the spectrum, at $n=22,23$, the brief exponential behavior occurs. This behavior is again suggestive of the transition to exponential growth of the OTOC, as expected.

\begin{figure*}[!ht]
    \centering
    \textbf{Lennard-Jones \qquad\qquad\qquad\qquad\qquad\qquad\qquad\qquad} \textbf{Rb$_2$ $X^{1}\Sigma^+_g$}\par\medskip
    \includegraphics[scale=0.3,trim=15 10 15 0,clip]{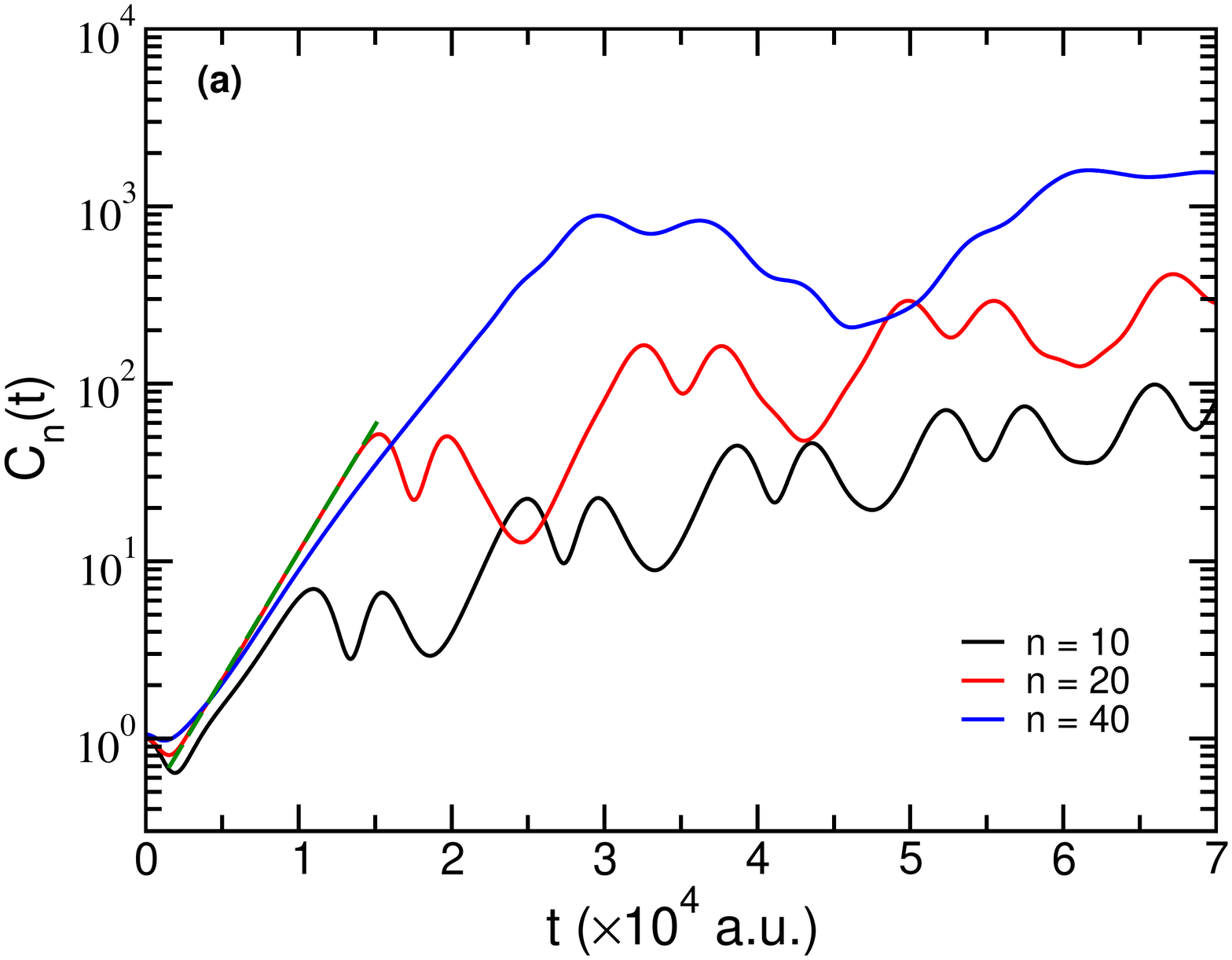}
    \includegraphics[scale=0.3,trim=15 10 15 0,clip]{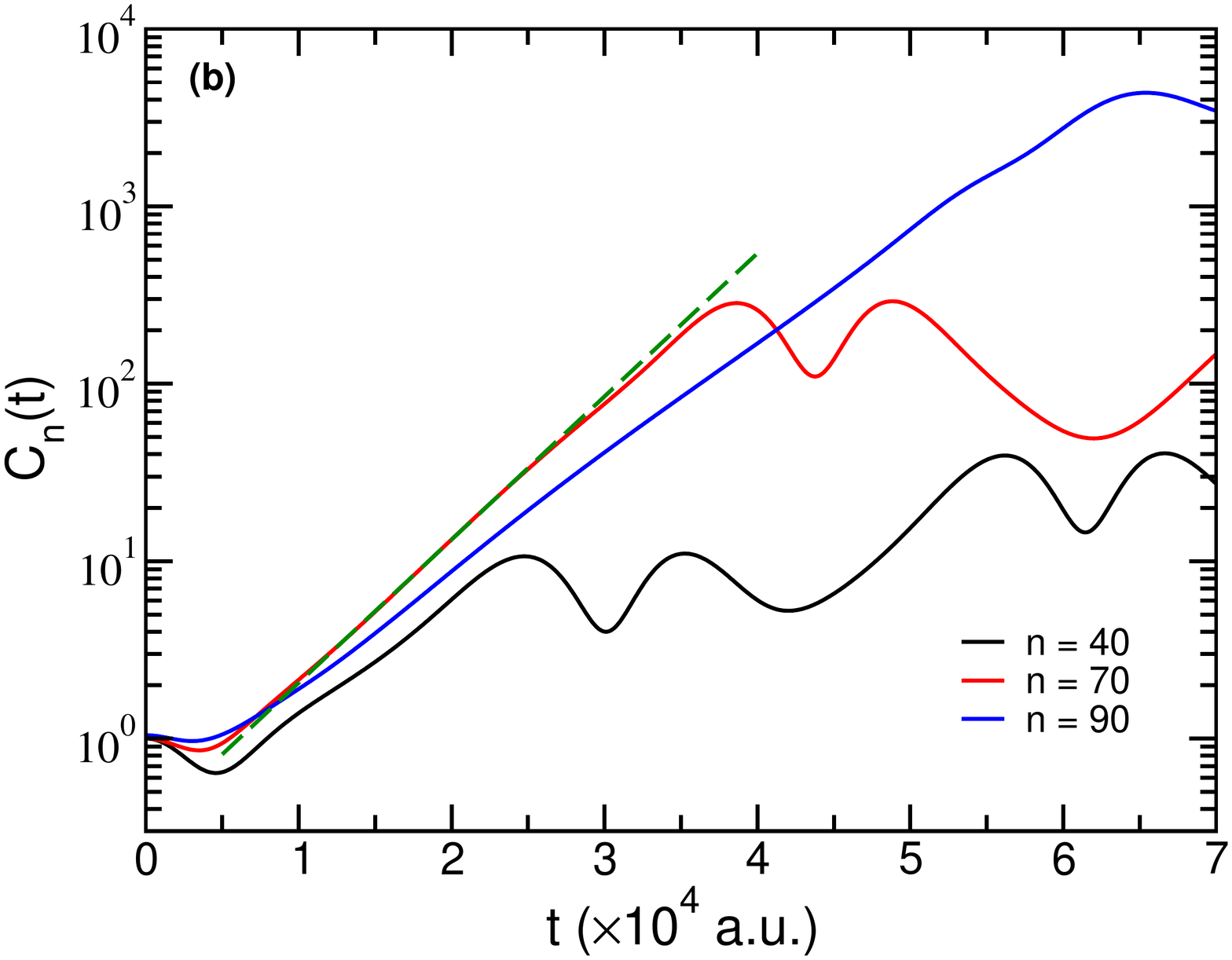}
    \caption{OTOCs of high energy modes. The dashed green lines are fitted with the exponential expression as discussed in the context.}
    \label{fig:m2}
\end{figure*}

We compute the OTOCs as function of $t$ for each energy level $n$ in the high energy modes which better exhibit exponential growth, at least for a short time. This behavior is shown for several values of $n$ in  Fig.~\ref{fig:m2}. For the states shown, there is a clear region where the OTOC grows exponentially, and the time interval over which the exponential growth occurs is different for each eigenstate $n$. For each potential, a representative exponential fit is shown as a green dashed line, fitting the OTOC for the $n=20$ level of the Lennard-Jones potential, and the $n=70$ level of the rubidium dimer potential.

\begin{figure*}[!ht]
    \centering
\textbf{Lennard-Jones \qquad\qquad\qquad\qquad\qquad\qquad\qquad\qquad} \textbf{Rb$_2$ $X^{1}\Sigma^+_g$}\par\medskip
    \includegraphics[scale=0.3,trim=15 10 15 0,clip]{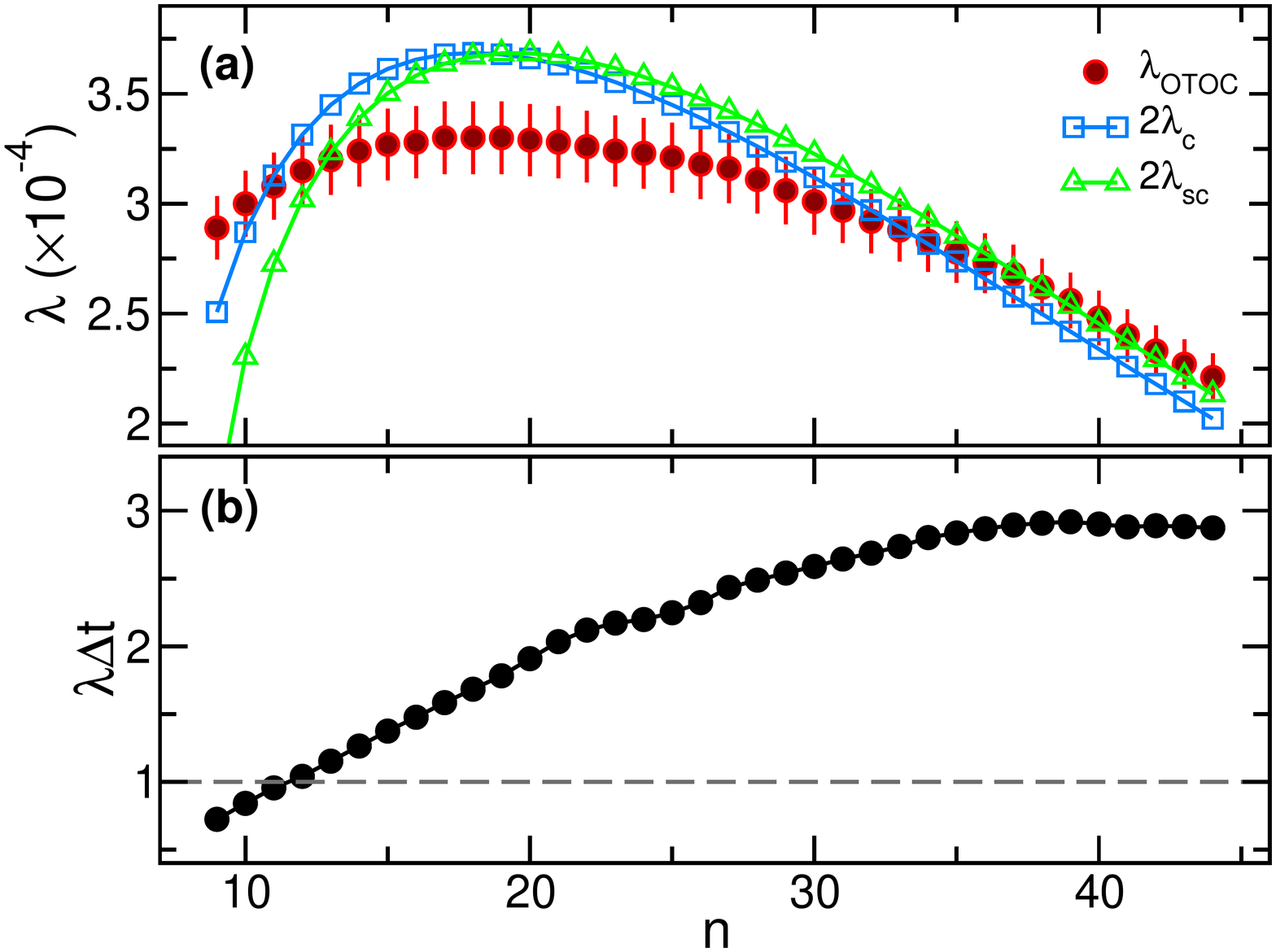}
 \includegraphics[scale=0.3,trim=15 10 15 0,clip]{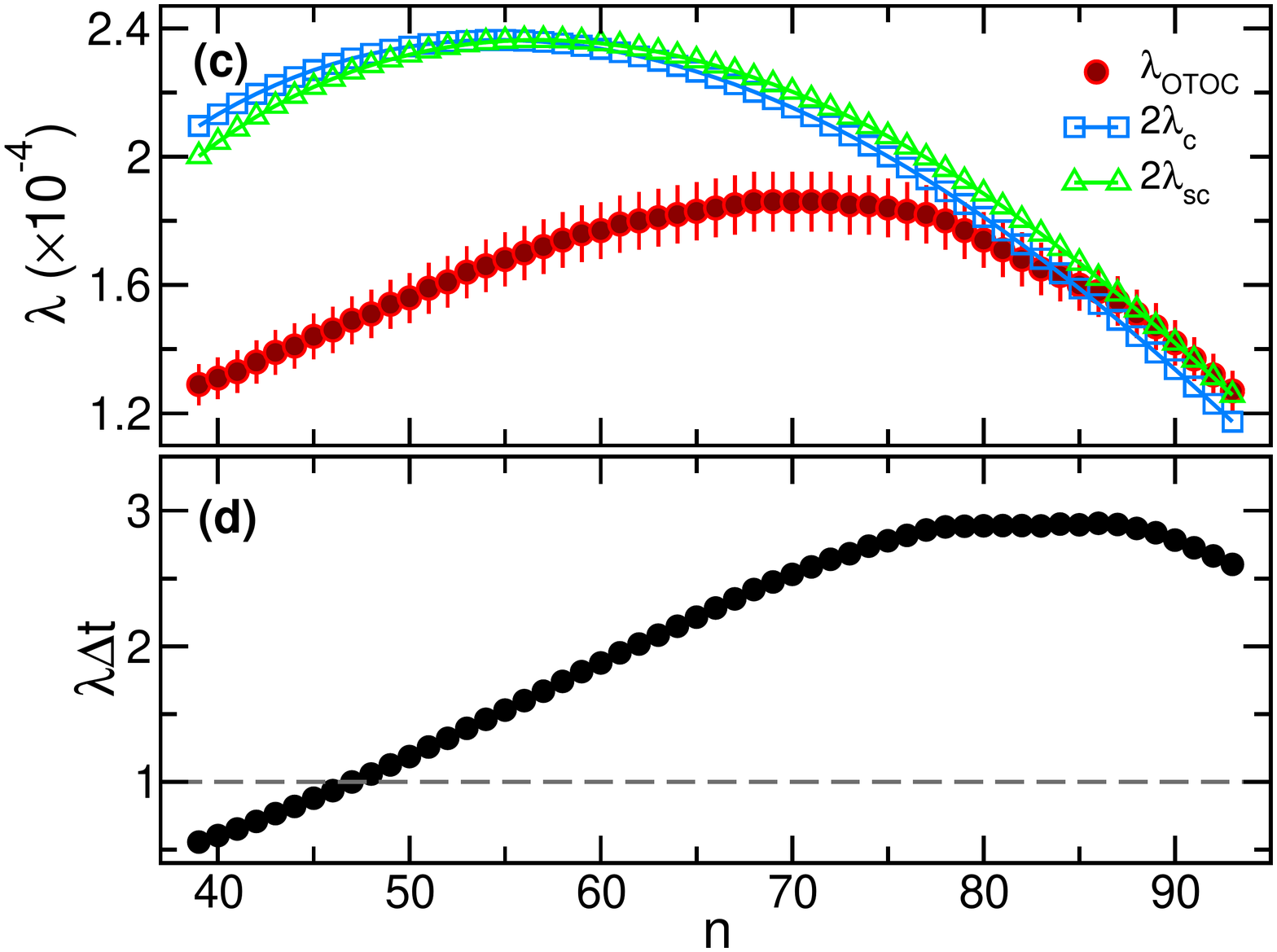} 
    \caption{Panels (a) and (c), the $n$-dependence of the quantum sensitivity parameters $\lambda_{\rm{OTOC}}$. The red solid dots are the fit values, whereas the cyan squares are the anti-harmonic approximation at the out classical turning points, and the blue triangles are the anti-harmonic approximation at the wavefunctions maximal. Panels (b) and (d) monitor the confidence of the fit process, $\Delta t$ is the fitting time interval.}
    \label{fig:fit}
\end{figure*}

Fitting each such curve to the form $\alpha \exp( \lambda_\mathrm{OTOC}\; t)$, defines an $n$-dependent sensitivity parameter $\lambda_{\mathrm{OTOC}}$. We denote this quantity as a ``sensitivity parameter'', rather than a Lyapunov exponent, recognizing that these one-dimensional systems are integrable. The sensitivity parameters extracted from this kind of exponential fit, for a number of different eigenstates $n$, are shown as the red dots in Figs.~\ref{fig:fit}(a) and (c), for the two potentials. The error bars represent a $95\%$ confidence interval for the fits. The parameter $\lambda_\mathrm{OTOC}$ shows a definite trend, first rising, then falling, as a function of $n$ in this part of the spectrum. The green and blue traces represent simple approximations, given in the next section.

An essential feature in any exponential growth trend, is that the growth occurs for a sufficient time to identify it as exponential. In the present case, each OTOC grows exponentially over some time interval $\Delta t$, over which time the fit is constructed. If this time is short compared to the derived exponential time constant, then it is difficult to say the behavior is truly exponential. Therefore, we track the product $\lambda_\mathrm{OTOC} \Delta t$ versus $n$, in Figs.~\ref{fig:fit}(b) and (d). If this product is greater than unity, then the exponential growth endures for at least one time constant. Over most of the spectrum shown, the product is indeed larger than one, meaning that the exponential growth is meaningful. This condition breaks down near the transition region, $n=6, 7$ for the Lennard-Jones potential, and $n=22, 23$ for the rubidium dimer $X^{1}\Sigma^+_g$ potential, where the transition to exponential growth is incomplete.

\section{Interpretation}
\label{sc:int}

As we can see in Fig.~\ref{fig:m1}, the OTOC is reasonably oscillatory for low energy levels of the potentials we have considered, which are approximately harmonic. It is only higher in the spectrum that $C_n(t)$ can grow exponentially as shown in Fig.~\ref{fig:m2}.  Concomitant with this, the higher energy levels have wave functions with large amplitude at the outer classical turning point $r_c$. Since the OTOC is an average over the wave function, it therefore arises rather locally from points near $r_c$. This is not unreasonable, as the sensitive dependence on initial condition can surely depend on what that initial condition actually is.

\subsection{Classical Sensitivity}

In any event, this localization in $r$ allows a closer comparison to be made between the classical and quantum mechanical versions.  For example, the classical motion may as well be something that starts at $r = r_0$, close to $r_c$, and remains close to that point.  In the neighborhood of $r_c$, we expand the potential to quadratic order,
\begin{align}
V(r) &\approx V(r_c) + V^{\prime}(r_c)(r - r_c) + \frac{ 1 }{ 2 }V^{\prime \prime}(r_c)(r - r_c)^2  \nonumber \\
&= \frac{ 1 }{ 2 } V^{\prime \prime}(r_c)(r - r_d)^2,
\end{align}
where in the second line we ignore an overall constant, and define $r_d = r_c - V^{\prime}(r_c)/V^{\prime \prime}(r_c)$.  For initial condition $r_0$ in the vicinity of $r_c$, the solution to the equation of motion $m{\ddot r} = - \partial V(r) / \partial r$ is
\begin{widetext}
    \begin{align}
r(t) &= r_d + (r_0 - r_d) \cos \left( \sqrt{ \frac{ V^{\prime \prime}(r_c) }{ \mu } } t \right)
 + \frac{ p_0 }{ \sqrt{ \mu V^{\prime \prime}(r_c)  }  }  \sin \left( \sqrt{ \frac{ V^{\prime \prime}(r_c) }{ \mu } } t \right) 
\end{align}
if $V^{\prime \prime}(r_c) >0$; and
\begin{align}
r(t) &= r_d + (r_0 - r_d) \cosh \left( \sqrt{ \frac{ -V^{\prime \prime}(r_c) }{ \mu } } t \right) 
+ \frac{ p_0 }{ \sqrt{ - \mu V^{\prime \prime}(r_c)  }  }  \sinh \left( \sqrt{ \frac{ -V^{\prime \prime}(r_c) }{ \mu } } t \right) 
\end{align}
if $V^{\prime \prime}(r_c) < 0$.
\end{widetext}
From these expressions it is clear that classical sensitivity to initial condition occurs, approximately, only when the potential has local negative curvature, leading to divergence $e^{\lambda_c t}$, in terms of a classical sensitivity parameter
 \begin{align} 
 \lambda_c = \sqrt{   \frac{ |V^{\prime \prime}(r_c)| }{  \mu } }.
 \end{align}

 \subsection{Quantum Sensitivity} 
 
A completely analogous situation holds in the quantum mechanical case.  Assuming a quadratic potential in the quantum Hamiltonian,
\begin{align}{\hat H} = \frac{ {\hat p}^2 }{ 2\mu } 
+ \frac{ 1 }{ 2 } V^{\prime \prime}(r_c)( {\hat r} - r_d )^2,
\end{align}
the Heisenberg position operator is evaluated using the Baker-Campbell-Hausdorff lemma, similar to what one does for the standard harmonic oscillator \cite{sakurai1995}.  For our present purposes, we consider only the case of negative curvature of the potential, where the operator becomes
\begin{widetext}
    \begin{align}
{\hat r}(t) &= r_d + ({\hat r}(0) - r_d) \cosh \left( \sqrt{ \frac{ -V^{\prime \prime}(r_c) }{ \mu } } t \right) 
+ \frac{ {\hat p}(0) }{ \sqrt{ - \mu V^{\prime \prime}(r_c)  }  }  \sinh \left( \sqrt{ \frac{ -V^{\prime \prime}(r_c) }{ \mu } } t \right) 
\end{align}
\end{widetext}
 Thus 
\begin{align}
\left[ {\hat r}(t), {\hat p}(0) \right] = i \cosh \left( \sqrt{ \frac{ -V^{\prime \prime}(r_c) }{ \mu } } t \right) 
\end{align}
Because the OTOC is defined in terms of the square of this quantity, its exponential growth in this approximation would be $\propto \left[ \exp( \sqrt{ V^{\prime \prime}(r_c) / \mu }t) \right]^2  = e^{2 \lambda_c t}$, that is, at a rate twice that of the classical exponential growth. The growth rate $2 \lambda_c$ is plotted in Fig.~\ref{fig:fit}.

A slightly better estimate of the OTOC's growth rate would evaluate the commutator, not at the classical turning point $r_c$, but at a nearby point $r_m$ where the wave function is maximal. To this end, we make a linear approximation of the potential near the turning point,
\begin{align}
V(r) \approx E + V^{\prime}(r_c)(r-r_c).
\end{align} 
Defining the characteristic length scale $\bar{r} = (\frac{1}{2\mu V^{\prime}(r_c)} )^{1/3}$, and in terms of the dimensionless variable $z = (r-r_c)/{\bar r}$, the resulting Schr$\ddot{\text{o}}$dinger equation becomes 
\begin{align}
\frac{ d^2 \psi }{ dz^2 } - z \psi = 0.
\end{align}
The solution to this is the usual Airy function, which, to a good approximation, is maximal at $z=-1$, whereby our bound state wave functions are maximal at $r=r_m$, where
\begin{align}
r_m = r_c - {\bar r} = r_c - \left( \frac{ 1 }{ 2 \mu V^{\prime}(r_c) } \right)^{1/3}.
\end{align}
Using this result defines a corrected, semi-classical version of the OTOC growth rate $2\lambda_{sc}$, where
\begin{align}
\lambda_{sc} = \sqrt{ \frac{ |V^{\prime \prime}(r_m)| }{ \mu } }.
\end{align}
This quantity, of course, takes a different value for each bound state $n$. Results of this approximation are also plotted in Fig.~\ref{fig:fit}. The agreement with the quantum OTOC is slightly improved, especially at high $n$.

\section{Conclusions and outlooks}
\label{sc:co}

We have verified, in realistic molecular potentials, a trend that has been seen elsewhere, namely, that exponential growth of the classical sensitivity and the quantum mechanical OTOC are linked, even in the absence of chaotic behavior.  The distinction is that in the molecular system we consider, the characteristic long-range van der Waals tail of these potentials produces a fairly explicit transition between regular and sensitive behavior of the dynamics as a function of the spectral level considered. Molecules are thus revealed as objects of variable sensitivity by this measure.

The lessons learned from this work should inform the investigation of far more complex molecules, in particular the ultracold lanthanide dimers alluded to at the outset.  Such molecules are expected to be truly chaotic, rather than merely sensitive, as the models presented here were.  Thus for example, Dy$_2$ molecules just below their dissociation threshold should exhibit exponential growth of their OTOCs in some way distinct from the sensitivity we have described, which may lead to additional insight into their dynamics.  This dynamics has the advantage that the OTOC will depend not only on the level considered, but also on the magnetic field applied in the laboratory, which is suspected to increase the degree of chaos in the molecules \cite{makrides2018}.

Looking ahead, we note that the exponential growth of the OTOC at relatively short times is not the only characteristic designating chaotic behavior. Indeed, quantum chaotic systems have been identified for which the quantum mechanical OTOC does not appear to grow exponentially \cite{bertini2018, markovic2022}. Nevertheless, the long-time behavior was shown to oscillate in non-chaotic systems, but to saturate in chaotic systems \cite{garcia2018, Bergamasco2019, kidd2021, markovic2022}. Though we have  not emphasized this, in the present integrable Lennard-Jones and rubidium dimer examples, the OTOC indeed oscillates at long times.  It is expected that, instead, saturation will occur for the case of Dy$_2$.

\section*{Acknowledgements}
%\nocite{*}
This material is based upon work supported by the National Science Foundation under Grants No. PHY 1734006 and PHY 2110327.
\bibliography{main}% Produces the bibliography via BibTeX.

\end{document}